\pdfoutput=1
\documentclass[11pt,twoside,a4paper,cmspaper,final,collab]{cms-tdr}
\def\svnVersion{e7c312f-D}\def\svnDate{2021/06/28}\def\cmsCernNoTag{CERN-EP-2021-008}\def\cmsCernDate{\today}\def\cmsMessage{"Published in Physical Review Letters as \href{http://dx.doi.org/10.1103/PhysRevLett.126.252002}{\doi{10.1103/PhysRevLett.126.252002}.}"}
\begin{document}\cmsNoteHeader{SMP-19-002}

\newcommand{\pphi}{\ensuremath{p_{\phi}}\xspace}
\newcommand{\pphimiss}{\ensuremath{\pphi^\text{miss}}\xspace}
\ifthenelse{\boolean{cms@external}}{\providecommand{\cmsLeft}{upper\xspace}}{\providecommand{\cmsLeft}{left\xspace}}
\ifthenelse{\boolean{cms@external}}{\providecommand{\cmsRight}{lower\xspace}}{\providecommand{\cmsRight}{right\xspace}}

\newlength\cmsTabSkip\setlength{\cmsTabSkip}{1ex}

\cmsNoteHeader{SMP-19-002}
\title{Measurement of the \texorpdfstring{$\PW\gamma$}{Wgamma} production cross section in proton-proton collisions at \texorpdfstring{$\sqrt{s} = 13\TeV$}{sqrt(s) = 13 TeV} and constraints on effective field theory coefficients}
 
\date{\today}

\abstract{A fiducial cross section for $\PW\gamma$ production in proton-proton collisions is measured at a center-of-mass energy of 13\TeV in 137\fbinv of data collected using the CMS detector at the LHC. The $\PW \to \Pe\nu$ and $\mu\nu$ decay modes are used in a maximum-likelihood fit to the lepton-photon invariant mass distribution to extract the combined cross section. The measured cross section is compared with theoretical expectations at next-to-leading order in quantum chromodynamics. In addition, $95\%$ confidence level intervals are reported for anomalous triple-gauge couplings within the framework of effective field theory.}

\hypersetup{
pdfauthor={CMS Collaboration},
pdftitle={Measurement of the  W-gamma production cross section in proton-proton collisions at sqrt(s) = 13 TeV and constraints on effective field theory coefficients},
pdfsubject={CMS},
pdfkeywords={CMS, standard model, multiboson, diboson}}

\maketitle

The associated production of a $\PW$ boson and a photon in proton-proton ($\Pp \Pp$) collisions corresponds to a fundamental process that has bearing on the basic ingredients of the standard model (SM). A precise measurement of the $\Pp \Pp \to \PW \Pgg$ cross section probes the $\PW\PW\Pgg$ triple-gauge coupling (TGC) and higher-order corrections to it. The structure and strength of the $\PW\PW\Pgg$ TGC are closely related to the SU(2)$\times$U(1) gauge symmetry of the SM and the mechanism for its breaking, which can be altered through the presence of new physics with alternative symmetries or symmetry-breaking mechanisms, such as composite $\PW$ models~\cite{Cui_2009}. Physics at a high energy scale can be described in a generic way in the framework of effective field theory (EFT), and the $\Pp \Pp \to \PW \Pgg$ production cross section has direct implications for the lowest-dimension operators in the EFT expansion, including $\mathcal{O}_{WWW} = \mathrm{Tr}[W_{\mu\nu}W^{\nu\rho}W^{\mu}_{\rho}]$, which directly affects the $\PW\PW\Pgg$ TGC~\cite{DEGRANDE201321}. Previous measurements of $\PW\Pgg$ production from the LHC use the data collected in $2011$ at a center-of-mass energy of $7\TeV$~\cite{PhysRevD.89.092005,PhysRevD.87.112003}. Here we report the first measurement of the $\Pp \Pp \to \PW \Pgg$ cross section at 13\TeV based on data collected by the CMS experiment in $2016$--$2018$, corresponding to an integrated luminosity of 137 \fbinv.

\begin{figure}[!h]
\centering
\includegraphics[width=0.3\textwidth]{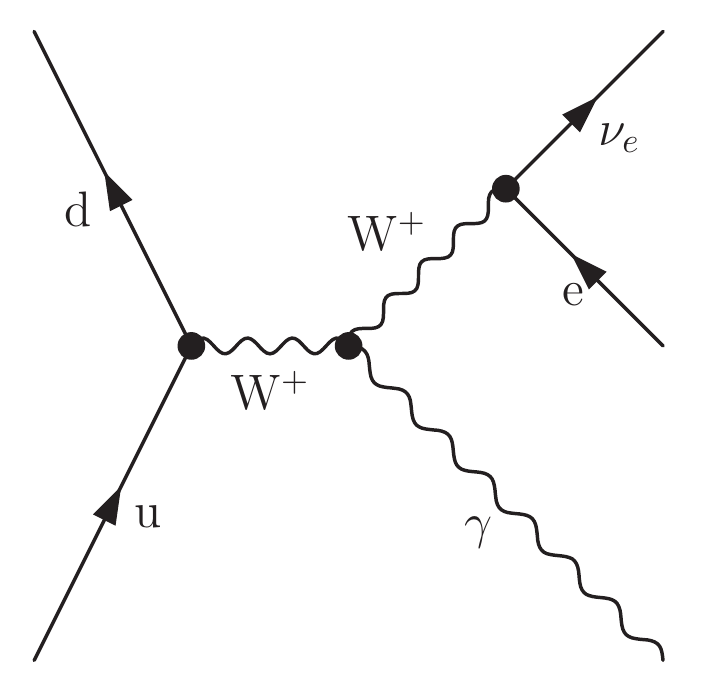}
\caption{Representative Feynman diagram for $\Pp\Pp \to \ell^+\nu_{\ell}\Pgg$ production with a TGC vertex.}
\label{fig:tgc_feynman_diagram}
\end{figure}

At leading order in quantum chromodynamics (QCD), $\ell^+\nu_{\ell}\Pgg$ and $\ell^-\overline{\nu}_{\ell}\Pgg$ (where $\ell = \Pe/\mu$) production in $\Pp\Pp$ collisions with an $s$-channel $\PW$ boson can proceed through initial-state radiation (ISR) from one of the incoming quarks, final-state radiation (FSR) from the outgoing charged lepton, or the $\PW\PW\Pgg$ TGC vertex shown in Fig.~\ref{fig:tgc_feynman_diagram}. At higher orders in QCD, additional quarks can appear in the final state, and the photon can arise by FSR from an outgoing quark or lepton. 

The central feature of the CMS apparatus is a superconducting solenoid of 6\unit{m} internal diameter, providing a magnetic field of 3.8\unit{T}. A silicon pixel and strip tracker, a lead tungstate crystal electromagnetic calorimeter (ECAL), and a brass and scintillator hadron calorimeter, each composed of a barrel and two end sections, are located within the magnetic field of the solenoid. Forward calorimeters extend the pseudorapidity ($\eta$) coverage provided by the barrel and end detectors. Muons are measured using gas-ionization chambers, including drift tubes, cathode strip chambers, and resistive plate chambers, embedded in the steel flux-return yoke outside the solenoid. A more detailed description of the CMS detector, as well as the definition of the coordinate system and the relevant kinematic variables, is reported in Ref.~\cite{Chatrchyan:2008zzk}.

Electrons and photons are measured in the range $\abs{\eta} < 2.5$ defined by the tracker acceptance. The energy of electrons is a combination of three measurements: the electron momentum at the primary interaction vertex as determined by the tracker~\cite{pvdefinition}, the energy of the corresponding ECAL cluster, and the energy sum of all bremsstrahlung photons spatially compatible with originating from the electron track. The photon momentum is determined solely using the energy measurement in the ECAL. The photon's ECAL cluster is required to be inconsistent with a charged-particle track reconstructed in the tracker~\cite{cmscollaboration2020electron}. Muons are measured in the pseudorapidity range $\abs{\eta} < 2.4$ and their momenta are determined using a global fit of muon measurements in the gas-ionization chambers and matched tracks in the silicon tracker~\cite{Sirunyan:2018}.

The missing transverse momentum vector $\ptvecmiss$ is computed as the negative vector \pt sum of all measured particles in an event, reconstructed with the particle flow algorithm~\cite{CMS-PRF-14-001}, and its magnitude is denoted by $\ptmiss$~\cite{Sirunyan_2019}. The $\ptvecmiss$ of an event is intended to represent the neutrinos associated with a single $\Pp \Pp$ interaction within a bunch crossing. The contribution to $\ptvecmiss$ due to particles from additional $\Pp \Pp$ interactions within the same bunch crossing (pileup) is mitigated through the pileup-per-particle identification algorithm~\cite{Bertolini:2014bba,cmspuppi}. The $\ptvecmiss$ is also modified to include corrections to the energy scale and resolution of the reconstructed jets in the event.

The $\PW\Pgg$ production cross section has been calculated with next-to-leading-order (NLO) QCD corrections at fixed order matched to a parton shower~\cite{Alwall2014,Barze2014}, with NLO electroweak corrections at fixed order~\cite{Denner2015}, and with next-to-next-to-leading-order (NNLO) QCD corrections at fixed order~\cite{Grazzini2018,Grazzini:2015nwa,PhysRevD.65.073003}. For an inclusive cross section, the NLO QCD corrections are large and positive, more than 100\% compared to the LO prediction, whereas the NLO electroweak corrections are negligible compared to experimental precision. The NNLO QCD corrections are positive and 20--30\% relative to the NLO QCD prediction.

The signal processes $\Pp \Pp \to \ell^+\nu_{\ell}\Pgg$ and $\Pp \Pp \to \ell^- \overline{\nu}_{\ell} \Pgg$ are simulated at NLO in QCD using \MGvATNLO version $5.2.6$~\cite{Alwall2014} with up to one jet in the matrix element calculation, merged with jets from the parton showering using the FxFx merging scheme \cite{Frederix2012}. These two processes are also simulated with \POWHEG version 2.0 using the C-NLO scheme~\cite{Nason:2004rx,Frixione:2007vw,Alioli:2010xd,Barze2014}, in which a QCD NLO accurate calculation is performed for up to one jet, and subsequently up to one additional jet or photon is emitted according to their respective Sudakov form factors. For both \MGvATNLO and \POWHEG, the parton showering and hadronization are performed using {\PYTHIA}8 version 8.226~\cite{Sjostrand:2014zea}, and the detector simulation is performed using \GEANTfour~\cite{AGOSTINELLI2003250}. To match data-taking conditions, we generate three sets of events corresponding to $2016$, $2017$, and $2018$. The {\PYTHIA}8 CUETP8M1~\cite{Khachatryan:2015pea} tune with the NNPDF30\_nlo\_nf\_5\_pdfas~\cite{Ball_2015} parton distribution functions (PDFs) are used for the $2016$ simulation, and the {\PYTHIA}8 CP5~\cite{Sirunyan:2019dfx} tune with the NNPDF31\_nnlo\_hessian\_pdfas~\cite{collaboration2017parton} PDFs are used for the $2017$ and $2018$ simulations.  The simulations include $\PW \to \Pgt \Pgngt$ decays, which are considered part of the signal when a $\Pgt$ decays with an emission of an electron or a muon. No EW or NNLO QCD corrections are applied.

We select $\PW^+\Pgg \to \ell^+\nu_{\ell}\Pgg$ and $\PW^-\Pgg \to \ell^-\overline{\nu}_{\ell}\Pgg$ events from the set of events that pass a level-one~\cite{Sirunyan:2020zal} and a high-level~\cite{Khachatryan:2016bia} trigger that require a single muon or electron that is isolated from other detector activity and, therefore, is likely to be promptly produced as opposed to produced during the hadronization of a jet. The $\pt$ threshold of the high-level trigger lepton varies between $24$ and $32\GeV$, depending on the year of data taking and the lepton flavor. We require the presence of a single high-quality~\cite{CMS:EGM-14-001} reconstructed photon, $\ptmiss$ exceeding $40\GeV$, and that the isolated electron or muon satisfies additional quality criteria~\cite{Sirunyan:2018,Khachatryan:2015hwa}. Offline kinematical requirements on the selected objects, based on the detector acceptance and the trigger thresholds, are photon $\pt > 25\GeV$, photon $\abs{\eta} < 2.5$, electron (muon) $\abs{\eta}< 2.5$ (2.4), electron (muon) $\pt > 30$ (26) or $>$ 35 (30)\GeV, depending on the year of data taking. To reduce the background from $\PZ\Pgg$ events, we reject events that contain an additional muon or electron with $\pt > 20\GeV$ that satisfies minimal quality criteria. Finally, $\Delta R = \sqrt{\smash[b]{\left(\Delta\eta\right)^2 + \left(\Delta\phi\right)^2}}$, where $\Delta \phi$ and $\Delta \eta$ are the spatial separations in azimuthal angle $\phi$ (in radians) and $\eta$ between the lepton and photon, is required to exceed $0.5$.

The signal is defined as the $\PW\Pgg$ process originating from a fiducial region defined with isolated prompt photons and isolated prompt dressed (as defined below) leptons. A lepton or photon is considered isolated if the \pt sum of all stable particles within $\Delta R = 0.4$, divided by the $\pt$ of the lepton or photon, is less than $0.5$. A lepton is considered prompt if it originates from the hard process; a photon is considered prompt if it originates from the hard process or an FSR or ISR process involving a particle that originates from the hard process. A lepton is dressed by adding to its four-momentum the four-momenta of all photons within $\Delta R = 0.1$; this procedure is intended to restore the lepton to its pre-FSR state. The fiducial region requirements are: photon and lepton $\abs{\eta}<2.5$ and $\pt > 25\GeV$, and $\Delta R($lepton$,$photon$) > 0.5$.

Background processes containing a prompt lepton and a prompt photon, including $\PZ\Pgg$ production, $\ttbar\Pgg$ production, and $VV\Pgg$ (where $V = \PW/\PZ$) production are simulated using \MGvATNLO and {\PYTHIA}8, in a manner similar to that for the signal samples. The background due to photon conversions in the detector material that lead to reconstructed electrons is estimated with a simulated sample of $\Pgg\Pgg$ events made with \SHERPA version 2.2.5~\cite{Gleisberg_2009}. The background due to events containing nonprompt leptons and photons, including those from instrumental mismeasurements and genuine leptons or photons within jets, is estimated from data. The ratio of well-isolated, high-quality leptons to less-well-isolated, lower-quality leptons is measured in a dijet control region in data as a function of the lepton $\abs{\eta}$ and $\pt$, and corrected for prompt leptons and prompt photon conversions based on simulated samples. A similar procedure is applied for photons based on a $\PW$+jets control region that excludes the signal region. In the nonprompt photon case, a fit to the width of the photon ECAL shower is used to determine the nonprompt photon fraction in the well-isolated, high-quality category, as described in Ref.~\cite{Sirunyan_2020}. The two procedures are combined in a way that avoids double counting to estimate the contribution from events containing both a nonprompt lepton and a nonprompt photon. The background contribution from events that contain a prompt lepton from the primary interaction and a prompt photon from a pileup interaction, mainly $\PW$+jets primary interaction events with $\Pgg$+jets pileup interaction events, is estimated using simulated samples. Finally, the background from electron-induced photons, occurring when an electron track is misreconstructed in the tracker or not properly matched to the corresponding ECAL cluster, is estimated using a fit to the $m_{\ell\gamma}$ distribution in data, which is sharply peaked because of the $\PZ$ resonance, with a template constructed from simulation.

\begin{figure}[!h]
\centering
\includegraphics[width=0.49\textwidth]{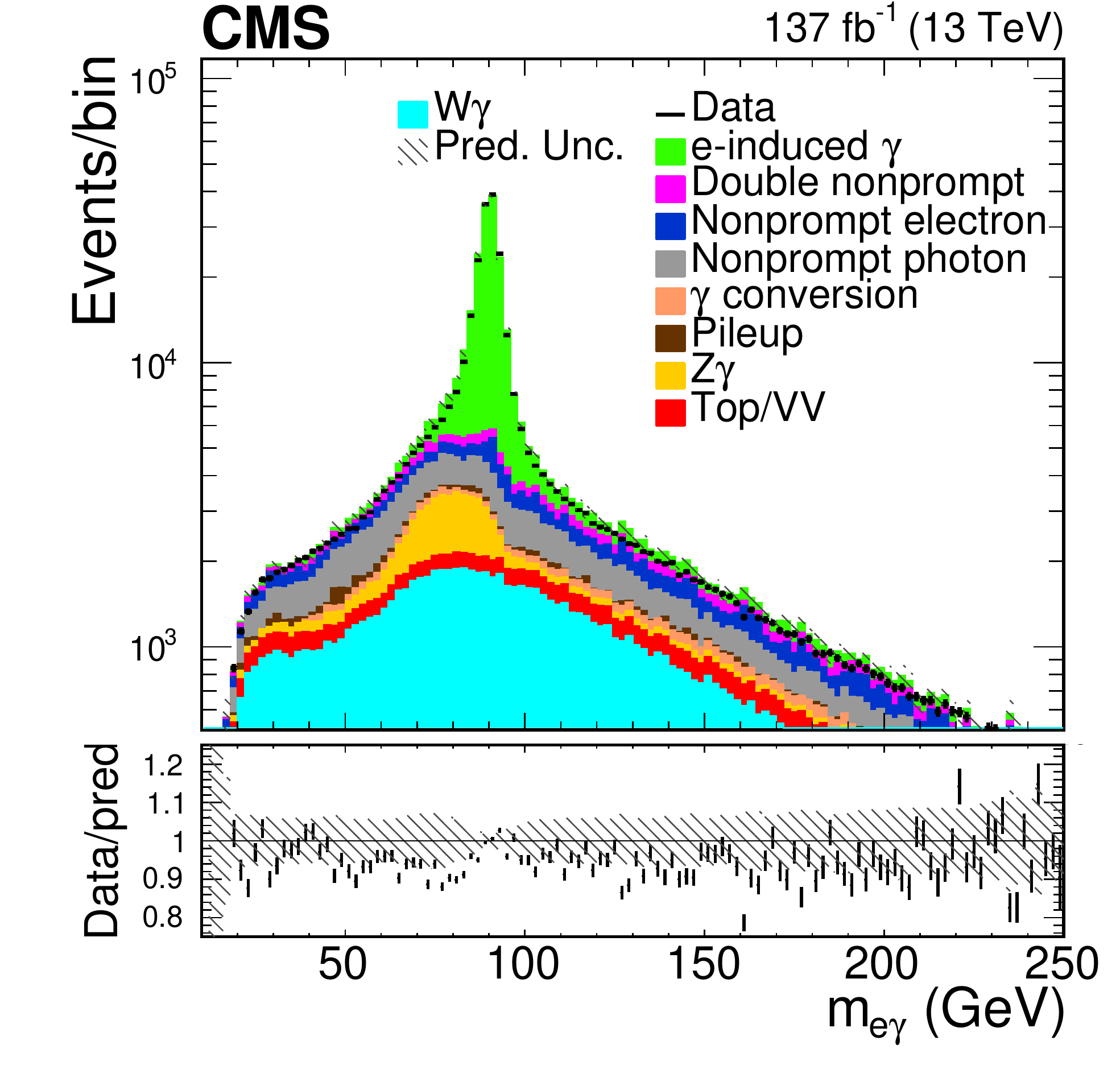} 
\includegraphics[width=0.49\textwidth]{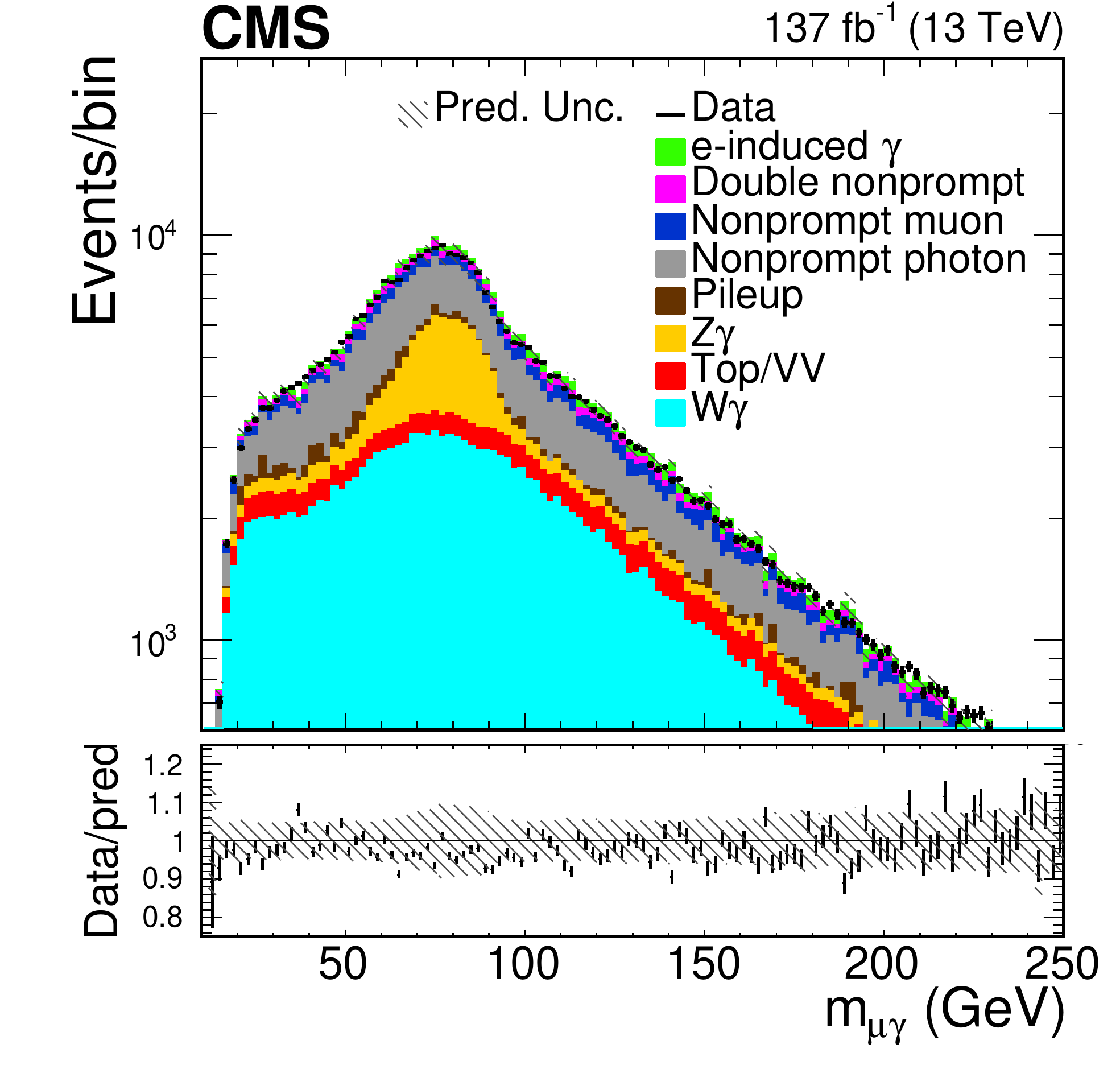}
\caption{Expected and observed distributions in the invariant mass of the lepton-photon system in the electron (\cmsLeft) and muon (\cmsRight) channels. The signal and background processes correspond to the estimates made before the fit, except that the normalization of the electron-induced photon (one of the free parameters) is scaled by $1.8$ from its prefit value. The uncertainty in the prediction (the hatched band) is the quadratic sum of the systematic uncertainties. The uncertainty in the data is statistical. The $\PW\Pgg$ label refers to the \MGvATNLO simulation of $\PW\Pgg$ events.}
\label{fig:observed_distributions}
\end{figure}

The observed distributions of $m_{\ell\gamma}$ are compared with the expected distributions based on the \MGvATNLO simulation in Fig.~\ref{fig:observed_distributions}. The experimental data agrees with the prediction within uncertainties. The expected and observed numbers of events are listed in Table~\ref{tab:n_event_yields}.

The signal strength is extracted from a binned maximum likelihood fit to the ${m_{\ell\Pgg}}$ distribution, where the likelihood function is the product of a Poisson probability density function for each bin. A simultaneous fit of the electron and muon channels is used for our main results; in addition, muon-channel-only and electron-channel-only fits are performed as a consistency check. In order to efficiently maximize the likelihood function with the large number of parameters that we consider, we use a \textsc{TensorFlow}-based minimizer~\cite{TensorFlow,CMS-PAS-SMP-18-012}. The fit is performed in the range $10$ to $250\GeV$ with $2\GeV$ bins. In the electron-channel-only fit and the simultaneous fit, the normalization of the electron-induced photon template is a free parameter in addition to the $\PW\Pgg$ normalization, whereas in the muon-channel-only fit, the normalization of the electron-induced photon template is constrained by a $100\%$ log-normal uncertainty around its nominal value and the $\PW\Pgg$ signal normalization is the only free parameter.

A variety of sources of systematic uncertainty are considered as nuisance parameters in the fit subject to log-normal constraints. Experimental sources of systematic uncertainty include: the jet energy scale and resolution (which affect the $\ptvecmiss$), the lepton and photon identification efficiencies, the pileup modeling, the integrated luminosity measurement, the statistical power of our simulated samples and data control regions, and the nonprompt photon and nonprompt lepton background estimation methods. Theoretical sources of systematic uncertainty include: the renormalization and factorization QCD scales, and PDFs. The renormalization and factorization QCD scales are varied by factors of $2$ and $1/2$, excluding the (2,1/2) and (1/2,2) cases, and the envelope of these variations is taken as the uncertainty. The systematic uncertainty due to the PDFs is calculated using the 32 additional members of the PDF4LHC15\_nnlo\_30\_pdfas PDF set following the PDF4LHC prescription for a Hessian PDF set~\cite{Butterworth_2016,Harland_Lang_2015,Ball_2015,PhysRevD.93.033006}. The uncertainties in the photon identification efficiency ($1$--$4$\%, depending on the photon \pt and $\eta$) and the integrated luminosity measurement (1.8\%) have the largest impact on the measurement.

The theoretical predictions of the cross section are $15.4 \pm 1.2 \text{ (scale)} \pm 0.1 \text{ (PDF)} \unit{pb}$ based on the NLO QCD \MGvATNLO simulation and $22.4 \pm 3.2 \text{ (scale)} \pm 0.1 \text{ (PDF)\unit{pb}}$ based on the NLO QCD \POWHEG simulation with the C-NLO scheme, where scale refers to QCD scale.

The measured cross section from the simultaneous fit with the uncertainties divided into statistical, experimental, and theoretical components is $\sigma = 15.58 \pm 0.75 \unit{pb} = 15.58 \pm 0.05 \stat \pm 0.73 \syst \pm 0.15 \thy \unit{pb}$. The measured cross section based only on the electron channel is $\sigma = 15.09 \pm 0.09 \stat \pm 1.02 \syst \pm 0.32 \thy \unit{pb}$ and the measured cross section based only on the muon channel is $\sigma = 15.77 \pm 0.06 \stat \pm 0.88 \syst \pm 0.12 \thy \unit{pb}$.

\begin{table}[htbp]
\topcaption{Expected and observed numbers of events. The signal and background yields correspond to the estimates made before the fit, except that normalization of the electron-induced photon yield (one of the free parameters) is scaled by $1.8$ from its pre-fit value. The uncertainty is the quadratic sum of the systematic uncertainties. The $\PW\Pgg$ label refers to the \MGvATNLO simulation of $\PW\Pgg$. The $\PW\Pgg$ signal and $\PW\Pgg$ nonfiducial are the contributions to the signal region from the $\PW\Pgg$ process originating from within and outside the fiducial region, respectively.}
\centering
\begin{scotch}{ccc}
Process  & $\Pe\gamma$ & $\mu\gamma$ \\
\hline
$\PW\gamma$ signal & $95\,953 \pm 6753$ & $164\,438 \pm 8\,773$ \\
$\PW\gamma$ nonfiducial & $1\,530 \pm 241$ & $2\,863 \pm 337$ \\
$Z\gamma$ & $22\,164 \pm 6\,173$ & $45\,227 \pm 11\,349$ \\
Top/$VV$ & $16\,501 \pm 879$ & $25\,517 \pm 952$ \\
Nonprompt photon & $46\,984 \pm 2\,249$ & $95\,838 \pm 4\,567$ \\
Nonprompt lepton & $27\,099 \pm 8\,169$ & $23\,008 \pm 6\,915$\\
Double nonprompt & $16\,264 \pm 4\,885$ & $14\,050 \pm 4\,219$ \\
$\Pe$-induced photon & $157\,209 \pm 42\,269$ & $14\,231 \pm 798$ \\ 
Pileup & $4\,892 \pm 475$ & $11\,085 \pm 782$ \\
Photon conversion & $8\,318 \pm 494$ & $0 \pm 0$ \\
[\cmsTabSkip]
Total & $396\,913 \pm 54\,686$ & $396\,257 \pm 22\,837$ \\
[\cmsTabSkip]
Observation & $385\,224$ & $395\,818$ \\
\end{scotch}
\label{tab:n_event_yields}
\end{table}

\begin{figure}[]
\centering
\includegraphics[width=0.49\textwidth]{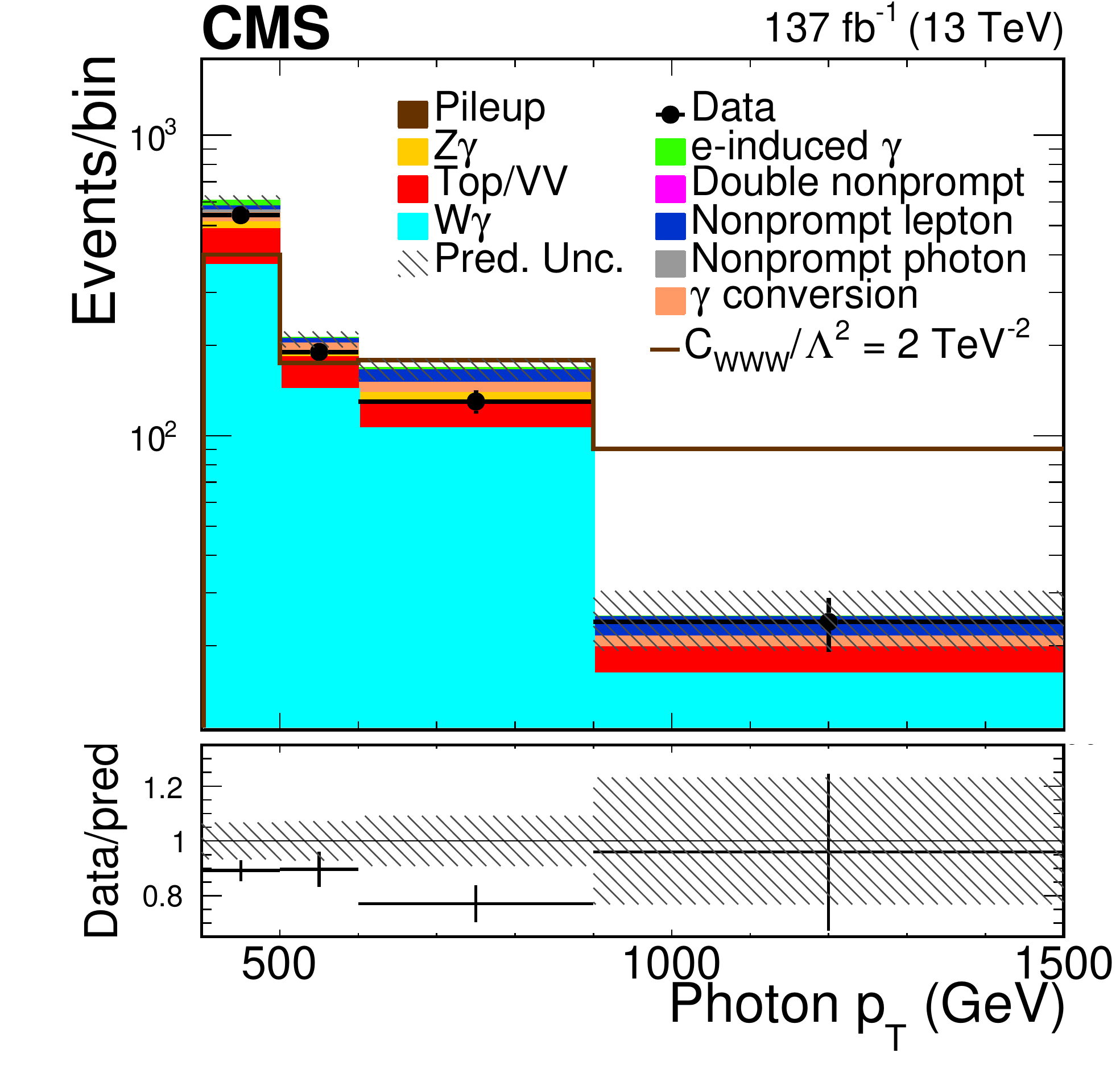}
\caption{The photon $\pt$ distribution used for the extraction of limits on dimension-six EFT operators. The expected yields correspond to the estimates made before the fit. The uncertainty in the prediction (the hatched band) is the quadratic sum of the systematic uncertainties. The uncertainty in the data is statistical. The last bin includes the overflow.}
\label{fig:ewdim6_photon_pt}
\end{figure}

Next, we search for new physics that could result in anomalous contributions to the cross section at high mass scale $\Lambda$. We consider an EFT in which dimension-six operators are added to the SM~\cite{DEGRANDE201321}
\begin{align*}
\mathcal{L}=\mathcal{L}_{\mathrm{SM}} + \sum_i\frac{c_i}{\Lambda^2}\mathcal{O}_{i}\mathrm{.}
\end{align*}
The operators that are relevant to $\PW\Pgg$ production are 
\begin{align*}
\mathcal{O}_{WWW} &= \mathrm{Tr}[W_{\mu\nu}W^{\nu\rho}W^{\mu}_{\rho}]\mathrm{,} \\
\mathcal{O}_{B} &= (D_{\mu}\Phi)^{\dagger}B^{\mu\nu}(D_{\nu}\Phi)\mathrm{,} \\
\mathcal{O}_{W\tilde{W}W} &= \mathrm{Tr}[\tilde{W}_{\mu\nu}W^{\nu\rho}W^{\mu}_{\rho}]\mathrm{, and} \\
\mathcal{O}_{\tilde{W}} &= (D_{\mu}\Phi)^{\dagger }\tilde{W}^{\mu\nu}(D_{\nu}\Phi)\mathrm{,}
\end{align*}
where $W^{\mu\nu}$ and $B^{\mu\nu}$ are the SU(2)$\times$U(1) field strength tensors, $\Phi$ is the Higgs field, and $\tilde{W}^{\mu\nu}$ is defined as $\epsilon^{\mu\nu\rho\sigma}W_{\rho\sigma}/2$ ($\epsilon^{\mu\nu\rho\sigma}$ is totally antisymmetric with $\epsilon^{0123} = 1$). The lowest dimension $CP$-even operator that directly alters the $\PW\PW\Pgg$ TGC is $\mathcal{O}_{WWW}$. The photon $\pt$ distribution shown in Fig.~\ref{fig:ewdim6_photon_pt} is used for the extraction of limits on the coefficients of these four operators. The NLO QCD reweighting feature of \MGvATNLO \cite{Mattelaer2016} is used to determine the yield of the $\PW\Pgg$ signal as a function of each operator coefficient. 

We compute expected and observed $95\%$ confidence level limits on each operator coefficient based on the profile likelihood ratio test statistic~\cite{CMS-NOTE-2011-005}. Each operator coefficient is scanned independently with all other operator coefficients set to zero. In addition to the sources of systematic uncertainty considered in the cross section fit, the $45\%$ difference between the \MGvATNLO and \POWHEG fiducial cross sections is assigned as an uncertainty in the normalization of the SM component of the model. The observed and expected limits are listed in Table~\ref{tab:aqgc_limits}. The observed limits on $c_{WWW} / \Lambda^2$ are decreased by a factor of $\approx$ $1.75$ relative to the previous best result~\cite{CMS_WV_atgc}. These limits can be converted through a linear relationship to limits on the parameters $\lambda_{\gamma}$, $\tilde{\lambda}_{\gamma}$, and $\tilde{\kappa}_{\gamma}$ in the Lagrangian approach to anomalous couplings, also known as the LEP parametrization, described in Ref.~\cite{DEGRANDE201321}. The expected limits on these parameters are $-0.0033 < \lambda_{\gamma} < 0.0033$, $-0.074 < \tilde{\kappa}_{\gamma} < 0.072$, and $-0.0016 < \tilde{\lambda}_{\gamma} < 0.0016$, while the corresponding observed limits are $-0.0035 < \lambda_{\gamma} < 0.0035$, $-0.066 < \tilde{\kappa}_{\gamma} < 0.065$, and $-0.0017 < \tilde{\lambda}_{\gamma} < 0.0017$.

\begin{table}[htbp]
\topcaption{Expected and observed 95\% confidence level limits on four dimension-six operator coefficients. The units of the limits are $\TeVns^{-2}$.}
\centering
\begin{scotch}{cccccc}
Coefficient & Exp. lower  & Exp. upper  & Obs. lower & Obs. upper   \\
\hline
$c_{WWW} / \Lambda^2$ & $-0.85$ & $0.87$ & $-0.90$ & $0.91$ \\
$c_{B} / \Lambda^2 $ & $-46$ & $45$ & $-40$ & $41$\\
$c_{\overline{W}WW} / \Lambda^2$ & $-0.43$ & $0.43$ & $-0.45$ & $0.45$ \\
$c_{\overline{W}} / \Lambda^2$ & $-23$ & $22$ & $-20$ & $20$ \\
\end{scotch}
\label{tab:aqgc_limits}
\end{table}

In summary, the cross section for $\Pp\Pp \to \PW\Pgg$ production has been measured at a center-of-mass energy of $13\TeV$ for the first time. The measured cross section in a defined fiducial region is $\sigma  = 15.58 \pm 0.05 \stat \pm 0.73 \syst \pm 0.15 \thy \unit{pb} = 15.58 \pm 0.75 \unit{pb}$, consistent with the \MGvATNLO next-to-leading-order (NLO) quantum chromodynamics (QCD) prediction of $\sigma = 15.4 \pm 1.2 \text{ (scale)} \pm 0.1 \text{ (PDF)} \unit{pb}$, and less than the \POWHEG NLO QCD prediction of $\sigma = 22.4 \pm 3.2 \text{ (scale)} \pm 0.1 \text{ (PDF)}\unit{pb}$. The cross sections in the electron and muon channels are consistent with each other. The high tail of the photon transverse momentum distribution is used to set 95\% confidence level limits on dimension-six effective field theory parameters, including the most stringent limit to date on the coefficient of $\mathcal{O}_{WWW}$, the lowest dimension $CP$-even operator that directly alters the $\PW\PW\Pgg$ TGC.

\begin{acknowledgments}
    We congratulate our colleagues in the CERN accelerator departments for the excellent performance of the LHC and thank the technical and administrative staffs at CERN and at other CMS institutes for their contributions to the success of the CMS effort. In addition, we gratefully acknowledge the computing centers and personnel of the Worldwide LHC Computing Grid and other centers for delivering so effectively the computing infrastructure essential to our analyses. Finally, we acknowledge the enduring support for the construction and operation of the LHC, the CMS detector, and the supporting computing infrastructure provided by the following funding agencies: BMBWF and FWF (Austria); FNRS and FWO (Belgium); CNPq, CAPES, FAPERJ, FAPERGS, and FAPESP (Brazil); MES (Bulgaria); CERN; CAS, MoST, and NSFC (China); COLCIENCIAS (Colombia); MSES and CSF (Croatia); RIF (Cyprus); SENESCYT (Ecuador); MoER, ERC PUT and ERDF (Estonia); Academy of Finland, MEC, and HIP (Finland); CEA and CNRS/IN2P3 (France); BMBF, DFG, and HGF (Germany); GSRT (Greece); NKFIA (Hungary); DAE and DST (India); IPM (Iran); SFI (Ireland); INFN (Italy); MSIP and NRF (Republic of Korea); MES (Latvia); LAS (Lithuania); MOE and UM (Malaysia); BUAP, CINVESTAV, CONACYT, LNS, SEP, and UASLP-FAI (Mexico); MOS (Montenegro); MBIE (New Zealand); PAEC (Pakistan); MSHE and NSC (Poland); FCT (Portugal); JINR (Dubna); MON, RosAtom, RAS, RFBR, and NRC KI (Russia); MESTD (Serbia); SEIDI, CPAN, PCTI, and FEDER (Spain); MOSTR (Sri Lanka); Swiss Funding Agencies (Switzerland); MST (Taipei); ThEPCenter, IPST, STAR, and NSTDA (Thailand); TUBITAK and TAEK (Turkey); NASU (Ukraine); STFC (United Kingdom); DOE and NSF (USA).
\end{acknowledgments}

\bibliography{auto_generated}
\cleardoublepage \appendix\section{The CMS Collaboration \label{app:collab}}\begin{sloppypar}\hyphenpenalty=5000\widowpenalty=500\clubpenalty=5000\vskip\cmsinstskip
\textbf{Yerevan Physics Institute, Yerevan, Armenia}\\*[0pt]
A.M.~Sirunyan$^{\textrm{\dag}}$, A.~Tumasyan
\vskip\cmsinstskip
\textbf{Institut f\"{u}r Hochenergiephysik, Wien, Austria}\\*[0pt]
W.~Adam, J.W.~Andrejkovic, T.~Bergauer, S.~Chatterjee, M.~Dragicevic, A.~Escalante~Del~Valle, R.~Fr\"{u}hwirth\cmsAuthorMark{1}, M.~Jeitler\cmsAuthorMark{1}, N.~Krammer, L.~Lechner, D.~Liko, I.~Mikulec, F.M.~Pitters, J.~Schieck\cmsAuthorMark{1}, R.~Sch\"{o}fbeck, M.~Spanring, S.~Templ, W.~Waltenberger, C.-E.~Wulz\cmsAuthorMark{1}
% ...
\vskip\cmsinstskip
\dag: Deceased\\
1:  Also at Vienna University of Technology, Vienna, Austria\\
% ...
\end{sloppypar}
\end{document}